\documentclass[pra,twocolumn,a4paper]{revtex4}
\usepackage[english]{babel}
\usepackage{graphicx,amsmath,bm, color}
\usepackage{ mathrsfs }
\usepackage{longtable}
\usepackage{hyperref}
\providecommand{\openone}{\leavevmode\hbox{\small1\kern-3.8pt\normalsize1}} %signe "identit"

\providecommand{\bra}[1]{\langle#1|}
\providecommand{\ket}[1]{|#1\rangle}

\providecommand{\ketbra}[2]{|#1\rangle\kern-2.8pt\langle#2|}

\usepackage{chngcntr}
\def\bra#1{\mathinner{\langle{#1}|}}
\def\ket#1{\mathinner{|{#1}\rangle}}

\def\text#1{\textrm{#1}}

\counterwithin{paragraph}{section}

\newcommand{\Tr}[1]{\mathrm{Tr}\!\left(#1\right)}

\newcommand{\be}{\begin{equation}}
\newcommand{\ee}{\end{equation}}

\renewcommand{\arraystretch}{1.5}

\begin{document}

\title{Bipartite nonlocality with a many-body system}

\author{Enky Oudot}
\affiliation{Quantum Optics Theory Group, University of Basel, CH-4056 Basel, Switzerland}
\author{Jean-Daniel Bancal}
\affiliation{Quantum Optics Theory Group, University of Basel, CH-4056 Basel, Switzerland}
\author{Pavel Sekatski}
\affiliation{Quantum Optics Theory Group, University of Basel, CH-4056 Basel, Switzerland}
\author{Nicolas Sangouard}
\affiliation{Quantum Optics Theory Group, University of Basel, CH-4056 Basel, Switzerland}

\date{\today}

\begin{abstract}
We consider a bipartite scenario where two parties hold ensembles of $1/2$-spins which can only be measured collectively. We give numerical arguments supporting the conjecture that in this scenario no Bell inequality can be violated for arbitrary numbers of spins if only first order moment observables are available. We then give a recipe to achieve a significant Bell violation with a split many-body system when this restriction is lifted. This highlights the strong requirements needed to detect bipartite quantum correlations in many-body systems device-independently.
\end{abstract}
\maketitle

\section{Introduction}
In a Bell test, distinct parties initially share a resource such as a quantum state. They are then given measurement settings that they use to obtain measurement outcomes. The joint statistics of their outcomes, conditioned on the settings, can then be used to reveal a number of properties. For instance, it can be shown that certain statistics are not compatible with pre-established agreements, as highlighted by the violation of a Bell inequality~\cite{Brunner14}. Recently, a theoretical demonstration showed that some many-body quantum states are able to violate multipartite Bell inequalities~\cite{Tura14, Wagner17, Tura17}. So far, no violation of these inequalities could be observed experimentally due to the challenge of addressing individual spins in a large ensemble. Nevertheless, witnesses were constructed and used to demonstrate the presence of Bell correlations in such states, i.e. their capacity to violate a Bell inequality~\cite{Schmied16,Engelsen17,Baccari18}. While such states are quite different from the quantum states often considered in the studies of Bell nonlocality, they are of particular relevance to many-body physics. It is then a natural question to ask whether a true Bell violation could be observed with such states. \\

Here, we make a step in this direction by considering the simple scenario in which hundreds of $1/2$-spins are split among just two protagonists, Alice and Bob. We then ask whether a Bell violation could be observed when the parties are restricted to perform collective measurements on their $1/2$-spin ensembles. Such a violation would provide a strong demonstration that mesoscopic systems can behave differently from the predictions of classical physics. Indeed, this conclusion would hold without the need to assume a quantum description of the setup. In particular, this conclusion would be independent of the Hilbert space dimension and of the proper calibration of the measurement device. For this reason, Bell tests performed on systems involving many spins are particularly appealing to show the limit of classical physics for describing mesoscopic systems. \\

\begin{figure}  
\includegraphics[width = 5cm]{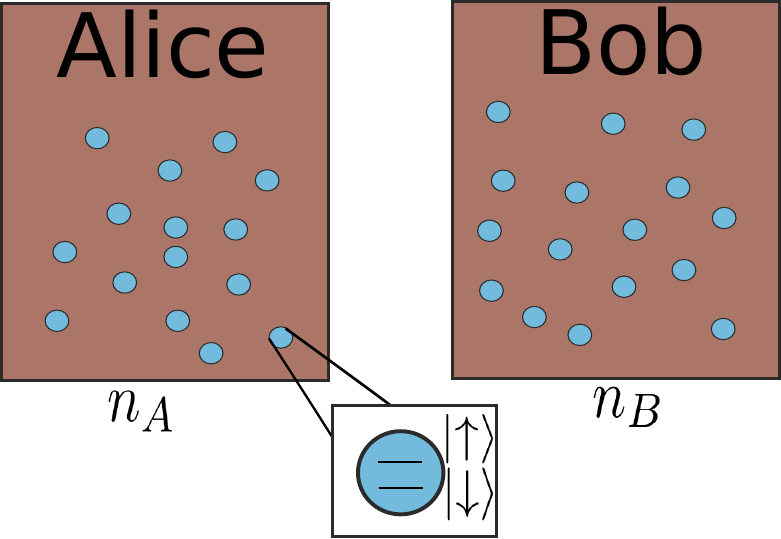}
\caption{Schematic representation of the scenario we consider, Alice and Bob have respectively  $n_A$ and $n_B$ indistinguishable spin $\frac{1}{2}$ and can measure them collectively with observables $\hat{J}_{\alpha}^A$ and $\hat{J}_{\beta}^B$  .}
\label{fig1}
\end{figure}

Bell tests involving collective measurements on many-body states could in principle be realized with a Bose-Einstein condensate. The basic idea is to use controlled interactions between the constituent bodies to create non-classical correlations between the internal states of these constituents \cite{Riedel10, Gross10}, the later being essentially $1/2$-spins. These spins could then be distributed \cite{Shin04, Schumm05} between Alice and Bob -- $n_A$ spins for Alice and $n_B$ for Bob -- before being measured, see Fig. \ref{fig1}. The experiment would then be repeated many times so that Alice and Bob can assess the expectation values of measurement results and demonstrate the Bell violation. In such a system, however, each protagonist can only measure his ensemble of particles collectively, that is, Alice can perform measurements of the form
\begin{equation}
\label{collectivespin_xA}
\hat{J}_{\alpha}^l= \frac{1}{2} \sum_{k=1}^{n_l} \vec{\alpha}.\vec{\sigma}_l^{(k)}, \-\ \text{with} \-\ l=A.\\
\end{equation}
$\vec{\sigma}_l^{(k)}$ is a vector having the 3 Pauli matrices as components and $\vec{\alpha}$ is a unit length vector with components $\alpha^x,$ $x=\{1,2,3\}$ fixing the measurement setting. The form of Bob's measurements, with $l=B$, is the same but the direction is labeled by $\vec{\beta}.$ The eigenvalues of $\hat{J}_{\alpha/\beta}^l$ are $-\frac{n_l}{2},...,\frac{n_l}{2}$ and correspond to the possible   measurement outcomes. The remaining question is what is the Bell inequality to be tested in such a scenario? We first focus on the simplest case where only first order moments of local collective spin observables, that is, $\langle \hat{J}_{\alpha}^A \rangle,$ $\langle \hat{J}_{\beta}^B \rangle$ and $\langle \hat{J}_{\alpha}^A\hat{J}_{\beta}^B\rangle$ can be evaluated. We then extend our analysis beyond this restriction.\\

While entanglement witnesses have been intensively studied in this scenario with first order moments~\cite{Durkin05, Toth04, Oudot17}, few results are known with respect to Bell nonlocality. One noticeable exception is Ref.~\cite{Ramanathan10} where authors showed that bipartite correlations issued from first order collective measurements can be reproduced by a local model if the number of measurement settings is smaller or equal to $n_A$ for Alice and $n_B$ for Bob. Beyond first order moments, it is worth noting that the use of collective measurements to violate a Bell inequality was also considered recently for a specific class of state with a strong tensor structure~\cite{Poh17, Zhou17}. The state that we are interested in here are however very different: they do not admit any tensor structure but rather are typically symmetric under spin exchange. \\

In section \ref{section_2}, we give numerical arguments suggesting that such a bipartite Bell violation may not be possible when considering only first order moments. We then show in section \ref{section_3} that a Bell inequality can be violated if each party performs parity measurements. We conclude in section \ref{section_4}.\\ 

\section{Attempts to find a bi-partite Bell correlation witness involving only first order moments of local collective spin components}
\label{section_2}
\subsection{Bell inequalities and Bell-correlation witnesses}
Let us focus on a Bell scenario in which Alice and Bob each have a measurement box with $m$ inputs $A_i$ and $B_j$ $i,j \in [1,\hdots,m]$ and $N$ spins locally, i.e. $N+1$ possible outcomes $-\frac{N}{2}, -\frac{N}{2}+1 \hdots,\frac{N}{2}.$ We consider Bell inequalities of the form
\begin{eqnarray}
\nonumber
&B_{N, m}^{\hat{w}, \vec{v}_a, \vec{v}_b}&=\sum_{i,j}^{m}w_{ij} \langle A_i B_j\rangle +\sum_{i}^{m}v_{a_{i}} \langle A_i \rangle +\sum_{j}^{m} v_{b_{j}} \langle B_j \rangle \\
\label{Bell_inequality}
&&\leq \ell_{N,m}^{\hat{w}, \vec{v}_a, \vec{v}_b}.
\end{eqnarray}
$\langle A_i \rangle$ for example is the expectation value of outcomes for Alice's measurement corresponding to the input $A_i.$ $w_{ij},$ $v_{a_{i}}$ and $v_{b_{j}}$ are weights, all taken in the interval \{-1,1\} without lost of generality.  $\ell_{N,m}^{\hat{w}, \vec{v}_a, \vec{v}_b}$ is the local bound, that is, the maximum value that the left hand side can take when considering bi-partite correlations steming from local models. This bound depends on the number of inputs and outcomes and on the weights, possibly in a non-trivial way, c.f. below. \\

Note that if we assign to each input an observable $A_i \rightarrow \hat{J}_{\alpha_{i}}^A$, one can associate to each instance of the inequality \eqref{Bell_inequality} a Bell operator
 \begin{equation}
 \label{Bell operator}
\mathscr{B}_{N,m}^{\hat{w}, \vec{v}_a, \vec{v}_b} =\sum_{i,j}^{m}w_{ij}\hat{J}_{\alpha_{i}}
^A\hat{J}_{\alpha_{j}}^B+\sum_{i}^{m}v_{a_{i}} \hat{J}_{\alpha_{i}}^A+\sum_{j}^{m}v_{b_j}\hat{J}_{\beta_{j}}^B
\end{equation}
so that a violation of 
\begin{equation}\label{Bell_operator_witness}
\langle \mathscr{B}_{N,m}^{\hat{w}, \vec{v}_a, \vec{v}_b} \rangle - \ell_{N,m}^{\hat{w}, \vec{v}_a, \vec{v}_b} \leq 0
\end{equation}
witesses bi-partite Bell correlations. Interestingly, the expression $\langle \mathscr{B}_{N,m}^{\hat{w}, \vec{v}_a, \vec{v}_b} \rangle$ might be simpler to evaluate experimentally than $B_{N, m}^{\hat{w}, \vec{v}_a, \vec{v}_b}.$ As an example, consider the simplest case $N=1,$ $m=2,$ $w_{ij}=(-1)^{i+j-1}$ and $v_{a_i} = v_{b_j} = 0$ $\forall i,j$ leading to the well known Clauser, Horne, Shimony, and Holt (CHSH) inequality \cite{CHSH69} for which the local bound is $2$, that is,   
\begin{equation}
\label{CHSH_inequality}
\langle A_1 B_1 \rangle + \langle A_2 B_1 \rangle + \langle A_1 B_2 \rangle - \langle A_2 B_2 \rangle \leq 2.
\end{equation}
Assigning $A_1$ to $\sigma_x,$ $A_2$ to $\sigma_z,$ and $B_{1/2}$ to $(\sigma_x \pm \sigma_z)\sqrt{2}$ which correspond to the setting choice maximizing the CHSH value for the singlet, we get the Bell-correlation witness
\begin{equation}
\label{CHSH_witness}
\langle \sigma_x \sigma_x \rangle + \langle \sigma_z \sigma_z \rangle \leq \sqrt{2}.
\end{equation}
The latter can be evaluated with two collective measurements while the former requires the assessment of four correlators. \\

Recently, such a reduction in the number of measurements, applicable when assuming that the measurements are known and trusted, was used to transform a Bell inequality involving an unbounded number of settings into a witness with only two global measurements~\cite{Wagner17}. It was also used to show that the Svetlichny inequality \cite{Svetlichny87}, which is a  Bell inequality for N parties requiring the measurement of $2^N$ correlators, reduces to a negativity condition on the mean value of an observable that can be evaluated with 2 measurement settings only \cite{Baccari18}. \\

There are thus at least two reasons for looking for Bell inequalities of the form~\eqref{Bell_inequality} despite the result from ~\cite{Ramanathan10} stating that no such inequality can be violated in our setting when the number of settings is less than the number of local spins, i.e. $m \leq N$. First, for every fixed value of $N$, there can exist a finite value of $m > N$ potentially allowing for a Bell violation. Second, in a scenario in which one would be happy to trust the quantum description of the measurements, a Bell inequality with $m >N$, although involving a large number of settings could lead to a witness for bipartite Bell correlations that would require a maximum of three measurements per party. Such a witness would allow for the detection of a strong form of quantum correlations in many-body systems. Compared to previously known Bell correlation witnesses which only demonstrate the presence of some form of Bell correlations among a large number of spins, this witness would demonstrate Bell correlations between two well-identified spin ensembles: between Alice's set of spins and Bob's. We show in the next two subsections how to compute the local and quantum bounds respectively. \\

\subsection{Local bound}
In order to determine the local bound $\ell_{N,m}^{\hat{w}, \vec{v}_a, \vec{v}_b}$ of the inequality \eqref{Bell_inequality}, one has to consider all possible local deterministic strategies, that is, strategies assigning locally an outcome taken from $\{-\frac{N}{2},-\frac{N}{2}+1,...,\frac{N}{2}\}$ for each of the $m$ settings. The local bound is simply the largest value that can be obtained from these deterministic strategies. On paper, it is sufficient to list all possible deterministic strategies to find the local bound. However, there are $(N+1)^{2m} $ deterministic strategies, which makes the computation of the local bound complicated even for small $N$ and $m$. \\

The number of relevant deterministic strategies is strongly reduced given the linearity of the inequality \eqref{Bell_inequality} with respect to the outcomes of Alice and Bob. To see this, let us consider an arbitrary deterministic strategy fixing the outcomes of Bob. The value of $B_{N, m}^{\hat{w}, \vec{v}_a, \vec{v}_b}$ is obtained from a quantity of the form
\begin{equation}
\label{local_bound}
\sum_{i}^{m}\alpha_{i}\langle A_i \rangle +C 
\end{equation}
where $\alpha_i = v_{a_{i}}+ \sum_{j} w_{ij} \langle B_j \rangle$ and $C=\sum_{j}^{m}v_{b_{j}} \langle B_j \rangle$. The maximum value of \eqref{local_bound} is  achieved for $A_i=\text{sign}(\alpha_{i})\frac{N}{2}$ and the same holds for Bob. This means that the optimal deterministic strategy is such that $A_i,B_j= \pm \frac{N}{2}$ $\forall i, j.$ Remarkably, this observation applies to all Bell inequalities of the form~\eqref{Bell_inequality}: the maximum local value of such inequalities can be obtained by strategies involving only the two extremal outcomes, for all coefficients $w_{ij},v_{a_i},v_{b_j}$. Conversely, it is possible to design an inequality of the form~\eqref{Bell_inequality} whose local bound is achieved by only one such strategy. This implies that the extremal vertices of the local polytope within the space parametrized by $\langle A_i\rangle, \langle B_j\rangle$ and $\langle A_i B_j\rangle$ are strategies assigning value $\pm \frac{N}{2}$ to all $A_i,B_j.$ This polytope is thus isomorphic to the bipartite Bell scenario with $m$ settings and $2$ possible outcomes. The number of extremal strategies for this polytope is $2^{2m}$, which is independent of the number of spins $N$. \\

In order to make the local bound independent of the number of spins we rescale the possible outcomes of $A_i$ and $B_j$ by $N$, defining $a_i=A_i/N$, $b_j=B_j/N$. This means that we now consider Bell inequalities of the form 
\begin{eqnarray}
\nonumber
&B_{N,m}'^{\hat{w}', \vec{v}'_a, \vec{v}'_b}&=\sum_{i,j}^{m} w_{ij}' \langle a_i b_j\rangle +\sum_{i}^{m} v_{a_{i}}' \langle a_i \rangle +\sum_{j}^{m} v_{b_{j}}' \langle b_j \rangle \\
\label{Bell_inequality_norm}
&&\leq \ell_{m}'^{\hat{w}', \vec{v}'_a, \vec{v}'_b},
\end{eqnarray}
where this time the $N+1$ possible outcomes of Alice and Bob take value in $-\frac12,-\frac12+\frac{1}{N},\ldots,\frac12$. For some parameters $\hat{w}', \vec{v}'_a, \vec{v}'_b$, the local bound $\ell_{m}'^{\hat{w}', \vec{v}'_a, \vec{v}'_b}$ of this Bell expression is achieved by considering the $2^{2m}$ local strategies for which $a_i,b_j=\pm\frac12$. This bound remains valid for all $N$. The local bound of the un-normalized Bell inequality~\eqref{Bell_inequality} is then given by $\ell_{N,m}^{\hat{w}, \vec{v}_a, \vec{v}_b} = \ell_{m}'^{N^2\hat{w}, N\vec{v}_a, N\vec{v}_b}$.\\

The Bell operator corresponding to this inequality can be built in terms of the re-normalized spin operators
\be 
\label{Nspin}
\hat{j}_{\alpha}^l = \frac{1}{N} \hat{J}_{\alpha}^l = \vec\alpha\cdot\vec s_l,
\ee 
with eigenvalues between $-\frac{1}{2}$ and $\frac{1}{2}$, where
\begin{equation}
\vec s_l = \frac{1}{2N} \sum_{k=1}^N \vec{\sigma}_l^{(k)}
\end{equation}
are the normalized spin projections:
\begin{equation}
\label{Bell operator_norm}
\mathscr{B}_{N, m}'^{\hat{w}', \vec{v}'_a, \vec{v}'_b} =\sum_{i,j}^{m} w_{ij}' \hat{j}_{\alpha_{i}}
^A \hat{j}_{\alpha_{j}}^B+\sum_{i}^{m}v_{a_{i}}' \hat{j}_{\alpha_{i}}^A+\sum_{j}^{m}v_{b_j}' \hat{j}_{\beta_{j}}^B.
\end{equation}
This operator is such that a violation of the inequality
\begin{equation}
\langle \mathscr{B}_{N, m}'^{\hat{w}', \vec{v}'_a, \vec{v}'_b} \rangle - \ell_{m}'^{\hat{w}', \vec{v}'_a, \vec{v}'_b} \leq 0
\end{equation}
witnesses Bell correlations. \\

We thus focus now on Bell inequalities of the form \eqref{Bell_inequality_norm}. This form has the advantage that the local bound is independent of $N$. Once the local bound of such a Bell inequality is found, we want to show that it is a non-trivial Bell inequality by checking that it can be violated, that is, it admits a quantum value larger than its local bound.\\

\subsection{Collective qubit bound} 

The maximal value of the inequality achievable by a quantum states of $N$ plus $N$ $\frac{1}{2}$-spins with local collective measurements $\max_{\rho_N} \Tr{\rho_N \mathscr{B}_{N, m}'^{\hat{w}', \vec{v}'_a, \vec{v}'_b}}$ can be obtained by finding the maximum eigenvalue of the operator $\mathscr{B}_{N, m}'^{\hat{w}', \vec{v}'_a, \vec{v}'_b}$ for all possible setting choice $\alpha_i, \beta_j.$ As we show below, the quantum bound $\langle \mathscr{B}_{N,m}'^{\hat{w}', \vec{v}'_a, \vec{v}'_b} \rangle$ decreases with $N$. Given this and the fact that the local bound is independent of $N$, we focus on the case $N=2$. Finding no Bell violation for $N=2$ and an arbitrary number of settings is sufficient to show that there is no non-trivial Bell inequality for any $N.$ We start by showing that indeed, the quantum bound $\langle \mathscr{B}_{m}^{\hat{w}, \vec{v}_a, \vec{v}_b} \rangle$ decreases with $N.$\\

The Bell operator $\mathscr{B'}_{N, m}^{\hat{w}', \vec{v}'_a, \vec{v}'_b}$ can be written in terms of the normalized spin projections as
\begin{equation}
\mathscr{B'}_{N, m}^{\hat{w}', \vec{v}'_a, \vec{v}'_b}= \vec{s}_A \cdot \hat{W} \cdot \vec{s}_B + \vec s_A \cdot\vec{V}_A + \vec s_B \cdot \vec{V}_B
\end{equation}
where $\hat{W}$ is a $3 \times 3$ matrix having elements $W^{xy}$ given by
\begin{equation}
W^{xy} = \sum_{i,j=1}^m w_{ij}' \alpha_i^x \beta_j^y,
\end{equation}
$\vec{V}_A$ is a vector with 3 components defined by $V_A^x = \sum_i v_{a_i}' \alpha_i^x$ and similarly for $\vec{V}_B.$\\

We now use $\vec k=(k_1, k_2, \ldots, k_M)$ with $1\leq k_i < k_{i+1}\leq N$ to denote a subset of $M<N$ spins and introduce the normalized spin observables over these spins
\begin{equation}
\vec u^{\vec k}_l = \frac{1}{2M}\sum_{i=1}^M \vec{\sigma}_l^{(k_i)}.
\end{equation}
This allows us to write our Bell operator for two sets $\vec k$ and $\vec k'$ of M spins as
\begin{eqnarray}
O^{\vec k, \vec k'}= \vec u^{\vec k} \cdot \hat{W} \cdot \vec u^{\vec k'} +\vec u^{\vec k} \cdot {\vec{V}_A} + \vec u^{\vec k'} \cdot {\vec{V}_B}.
\end{eqnarray}
Noticing that the spin projections for $M$ spins satisfy
\begin{equation}
\begin{split}
\sum_{\vec k} \vec u^{\vec k}_l &= \frac{1}{2M}\sum_{\vec k} \sum_{i=1}^M \vec{\sigma}_l^{(k_i)}\\
&=\frac{1}{2M}\binom{N-1}{M-1}\sum_{k=1}^N \vec{\sigma}_l^{(k)}\\
&=\binom{N}{M}\vec s_l
\end{split}
\end{equation}
where the sum on $\vec k$ runs over all choices of $M$ spins within the $N$ spins, we obtain
\begin{eqnarray}
\sum_{\vec k, \vec k'} O^{\vec k, \vec k'} = \binom{N}{M}^{2} \mathscr{B}_{N, m}'^{\hat{w}', \vec{v}'_a, \vec{v}'_b}.
\end{eqnarray}
Therefore, the maximum quantum value of the Bell operator for $M<N$ spins per side bounds the value of the Bell operator with $N$ spins:
\begin{equation}
\begin{split}
\underset{\rho_N}{\max}\ &\Tr{\rho_N \mathscr{B}_{N, m}'^{\hat{w}', \vec{v}'_a, \vec{v}'_b}} = \binom{N}{M}^{-2} \underset{\rho_N}{\max}\ \Tr{\rho_N \sum_{\vec k, \vec k'} O^{\vec k, \vec k'}}\\
&= \binom{N}{M}^{-2} \underset{\rho_N}{\max}\ \sum_{\vec k, \vec k'}  \Tr{\rho_N O^{\vec k, \vec k'}}\\
&\leq \binom{N}{M}^{-2} \sum_{\vec k, \vec k'} \underset{\rho_N}{\max}\ \Tr{\rho_N O^{\vec k, \vec k'}}\\
&= \underset{\rho_N}{\max}\ \Tr{\rho_N O^{\vec k, \vec k'}}\\
&= \underset{\rho_M}{\max}\ \Tr{\rho_M \mathscr{B}_{M, m}'^{\hat{w}', \vec{v}'_a, \vec{v}'_b}},\\
\end{split}
\end{equation}
where to go from the second to the third line, we let the optimization over the state be independent for each term in the sum. This shows that the maximal value of $\langle \mathscr{B}_{m}'^{\hat{w}', \vec{v}'_a, \vec{v}'_b} \rangle$ achievable with collective $1/2$-spin measurements can only decrease with the number of spins $N$.\\

\subsection{Numerical results}
Let us start this subsection by a summary of the two previous subsections: (i) The local bound of an inequality of the form \eqref{Bell_inequality_norm} is independent of the number of possible outcomes
(ii) Assuming collective measuremens on $\frac12$-spins, the quantum bound decreases while increasing the number of spins (or outcomes). Together, these two statements imply that if a Bell inequality of the form \eqref{Bell_inequality_norm} cannot be violated by performing collective measurements on an arbitrary state containing $N$ particles on each side, then it is also impossible to violate it by performing collective measurements on a state with more than $N$ particles on each side. Since the CHSH inequality is a non-trivial Bell inequality with $N=1$ spin locally, we focus on the case with $N=2$ spins locally. We know from Ref. \cite{Ramanathan10} that in this case, one needs at least 3 measurements settings locally to circumvent known local models and thus possibly violate a Bell inequality of the form \eqref{Bell_inequality_norm} with collective measurements.\\

In the case of 3 measurement settings, there is only one relevant Bell inequality with two outcomes~\cite{Froissart81}. Interpreting this inequality in terms of the normalized correlators of Eq.~\eqref{Bell_inequality_norm}, we easily compute the maximum quantum value that a state of 4 particles (2 at each location) can achieve for this inequality by optimizing the maximal eigenvalue of the corresponding Bell operator as a function of the measurement settings. We find that this inequality is not violated (up to the accuracy of the computation) with collective spin measurement of 2 particles at each side. This implies that this inequality is likely not to admit a violation with collective spin measurement irrespectively of the number of spins per side.\\
 
We now consider the case of 4 measurement settings. In this case, the full polytope with binary outcomes has been recently shown to contain 175 different orbits up to relabellings of parties, inputs and outputs~\cite{Deza16,VertesiLiang}. By gathering inequalities known in the litterature\cite{CHSH69,Froissart81,Collins04,Ito06,Avis05,Gisin07,OpenProblems,Brunner08,Pal09,Bancal10,Brunner09}, plus some other inequalities found by ourselves and by colleagues, we obtained a description of 175 such inequivalent families. They thus provide a full description of our polytope. We provide a list of these inequalities in the appendix. Focusing on these inequalities we computed the quantum bound as before with $N=2$ spins locally but with 4 measurements locally. We did not find any violation, which suggests that no inequality of the form \eqref{Bell_inequality_norm} with $m=4$ can be violated by collective spin observables.\\

For the case of 5 settings, the local polytope is not known even for the simplest case of binary outcomes. We thus proceed differently. This time, we sample different inequalities of the form \eqref{Bell_inequality_norm}, that is, we choose $\omega'_{ij},$ $v'_{a_i}$ and $v'_{b_j}$ at random and computed both the local bound and the quantum bound. Note that this time we are not restricting ourselves to binary outcomes. We test 400 000 Bell inequalities with $4\omega'_{ij},$ $2v'_{a_i}$ and $2v'_{b_j}$ distributed uniformly in $[-1,1]$. We do not find any violation. Repeating the same procedure for 6 settings locally, we again don't find any non-trivial Bell inequality. 6 settings is the maximum we succeeded to do because it becomes increasingly expensive to find the maximum quantum value. Also the parameter space increases with the number of settings, so one would require more and more random trials to span the space of inequalities when the number of settings increases. Still, altogether this suggests that none of the inequalities of the form \eqref{Bell_inequality_norm} can be violated by collective spin measurements when $N\geq 2$. \\

We have presented numerical arguments suggesting that it is not possible to violate an inequality of the form \eqref{Bell_inequality_norm} with collective spin measurements with 3, 4, 5 and 6 measurements settings whenever the number of spins locally is larger than two. In the next section, we show that the violation of a bipartite Bell inequality is possible when parity measurements are performed locally, for all spin number $N$. \\

\section{Proposal for the violation of a bi-partite Bell inequality with parity measurements}
\label{section_3}
\subsection{The scenario}
The scenario is similar to the one before. An ensemble of 1/2-spins is created in a quantum state before being shared between Alice and Bob. Each of them performs collective measurements of their spins $\hat{J}_\alpha^l$ but in opposition to the scenario of the previous section, there is no limit on the order of moments of collective spin components that can be measured. We assume in particular that Alice and Bob can assess precisely the parity of the measurement outcome at each run.  \\

Concretely, we consider an ensemble of $N$ $1/2$-spins encoded in the internal degree of atoms, that is, two atomic states 1 and 2. These spins are located in Alice's location. We thus call $\hat{a}_i$ and $\hat{a}_i^\dag$ with $i\in \{1,2\}$ the bosonic operators associated to each spin states so that the collective spin projections can be written as
\begin{eqnarray}
\label{collectivespin_xA}
 &&\hat{J}_x^A= \frac{1}{2}  (\hat{a}_1^{\dagger}\hat{a}_2+\hat{a}_1\hat{a}_2^{\dagger}),\\
\label{collectivespin_yA}
&&\hat{J}_y^A= \frac{1}{2 i} (\hat{a}_1^{\dagger}\hat{a}_2-\hat{a}_1\hat{a}_2^{\dagger}),\\
\label{collectivespin_zA}
&&  \hat{J}_z^A=\frac{1}{2}    (\hat{a}_1^{\dagger}\hat{a}_1-\hat{a}_2^{\dagger}\hat{a}_2).
\end{eqnarray}
We further consider that initially the spins point in the x direction and then undergoes one-axis twisting \cite{Ma11, Pezze16}. This results in a spin-squeezed state
\begin{equation}
\label{spinsqueezed_state}
\ket{\psi}=\frac{1}{\sqrt{N}} e^{-i\chi t (\hat{J}_z^A)^2} e^{-i\frac{\pi}{2} \hat{J}_y^A} \hat{a}_{1}^{\dag  N} \ket{0} 
\end{equation}
where $\ket{0}$ is the vacuum state for all modes.  The particles are then shared between Alice and Bob with a beam splitter type Hamiltonian, that is    
\begin{equation}
\label{split_state}
\ket{\phi}=e^{\frac{\pi}{4}(\hat{a}_1^\dag \hat{b}_1 + \hat{a}_2^\dag \hat{b}_2 - \text{h.c.})} \ket{\psi}.
\end{equation}
Here $\hat b_i$ and $\hat b_i^{\dagger}$ are bosonic operators for the spins located at Bob's location. In practice, each of these steps can be realized with a Bose-Einstein condensate where spin squeezing can be created using elastic collisions in state dependent potentials \cite{Riedel10, Gross10}. The spatial splitting can then be done by slowly raising a barrier in a state-independent potential as in Refs. \cite{Shin04, Schumm05}. 

\begin{figure} 
\includegraphics[width = 7.5cm]{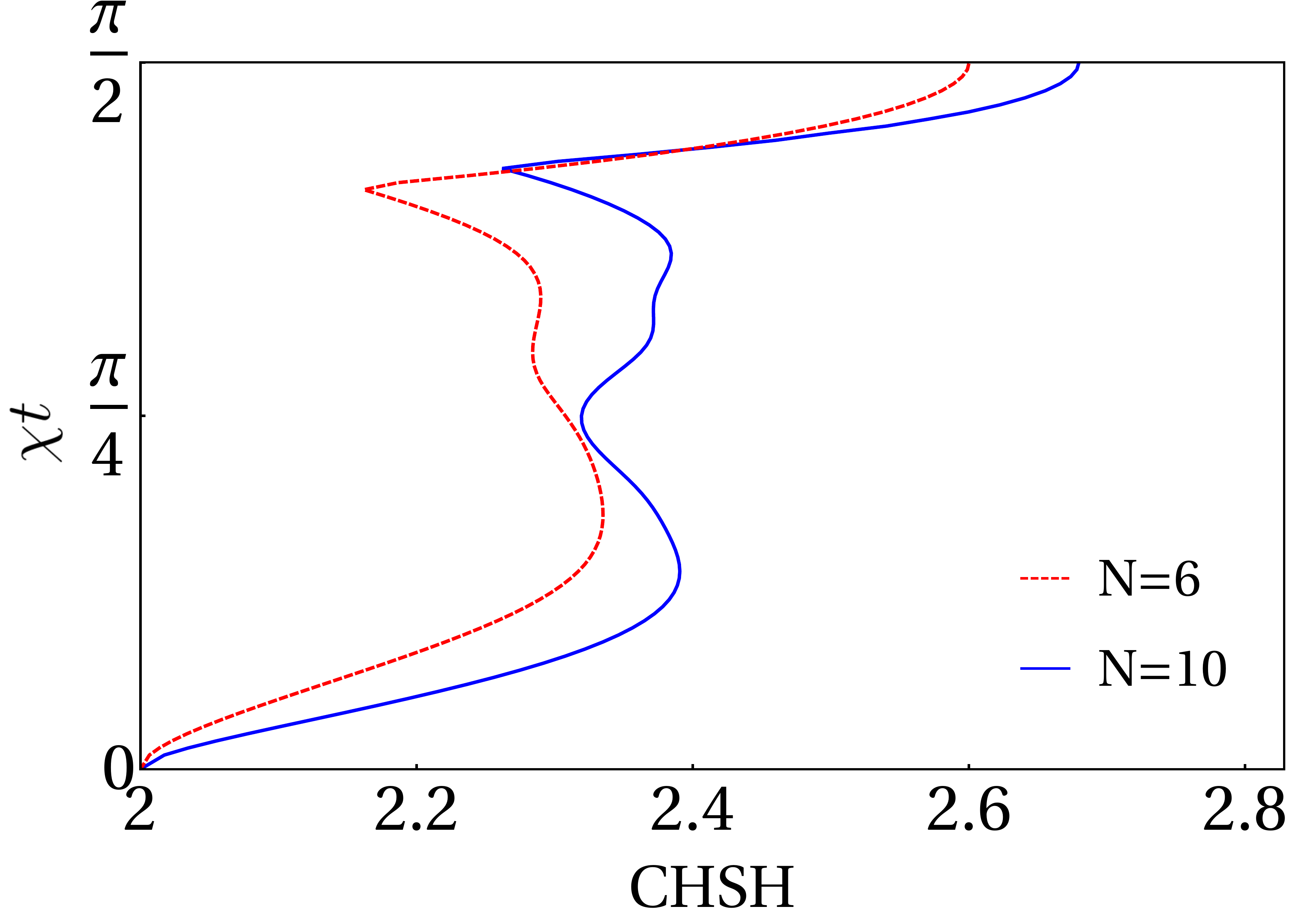}
\caption{Violation of inequality \eqref{CHSH_inequality} for a spin squeezed state $\ket{\phi}$ with 6 and 10 atoms as a function of the squeezing parameter $\chi t$}
\label{fig2}
\end{figure}
 
\subsection{Probability distribution} Alice and Bob are sharing the state $\ket{\phi}$ and want to compute the value of the CHSH quantity using measurements $\hat{J}_{\alpha}^A$ and $\hat{J}_{\beta}^B$ where each setting is specified by two angles $\{\theta_l,$ $\phi_l\}$ via $\vec{\alpha}=(\sin{\theta_\alpha}\cos{\phi_\alpha},\sin{\theta_\alpha}\sin{\phi_\alpha},\cos{\theta_\alpha})$ and similarly for $\vec \beta.$ Let $|m_a,n_a-m_a\rangle$ be the state of Alice with a total spin number $n_a$ and $m_a$ excitations in the state 2. Since $\hat{J}_{z}^A$ is half the population difference between the spin states 1 and 2, we have  $\hat{J}_{z}^A |m_a,n_a-m_a\rangle = \frac{1}{2}(n_a-2m_a)|m_a,n_a-m_a\rangle.$ Since any operators $\hat{J}_{\alpha}^A$ is linked to $\hat{J}_{z}^A$ by a unitary, we can express its eigenstates as a function of $|m_a,n_a-m_a\rangle$ through $\ket{m_a^\alpha,n_a-m_a^\alpha} = \sum_{k=-n_a/2}^{n_a/2}\mathcal{D}_{k,m_a}^{n_a}(\theta_a,\phi_a)\ket{\frac{n_a-2k}{2},\frac{n_a+2k}{2}}$ where $\mathcal{D}_{k,m_a}^{n_a}(\theta_a,\phi_a)$ is similar to a Wigner matrix:
\begin{equation}
\label{D k,m_a n_a}
e^{-i \phi_A k}\bra{\frac{n_a-2k}{2},\frac{n_a+2k}{2}}e^{-i \theta_A \hat{J}_y^A}\ket{m_a,n_a-m_a}.
\end{equation}
The same basis $\{|m_a,n_a-m_a\rangle\}$ can also be used to express the state that Alice and Bob share, that is $\ket{\phi}$ can be written as 
\begin{equation}
\label{phi}
\frac{1}{2^N}\sum_{m=0}^N\sum_{k=0}^m\sum_{l=0}^{N-m} C_{m,k,l}\ket{k,l}_A\ket{m-k,N-m-l}_B
\end{equation}
where $C_{m,k,l}=\sqrt{\binom {m} {N}\binom {k} {m}\binom {l} {N-m}}e^{-i\chi t (m-\frac{N}{2})^2}.$ When Alice and Bob measure $\hat{J}_{\alpha}^A$ and $\hat{J}_{\beta}^B$ respectively, the probability with which they find the eigenvalues $k_{\alpha}^a$ and $k_{\beta}^b$ is given by 
\begin{equation}
\begin{split}
P(k_{\alpha}^a,k_{\beta}^b|&\hat{J}_{\alpha}^A, \hat{J}_{\beta}^B)= \sum_{n_a=0}^N \left|\bra{\frac{n_a-2k_\alpha^a}{2},\frac{n_a+2k_\alpha^a}{2}}\right.\\
&
\left.\otimes\bra{\frac{N-n_a-2k_\beta^b}{2},\frac{N-n_a+2k_\beta^b}{2}}\phi\rangle\right|^2
\end{split}
\end{equation}
This probability can be efficiently calculated using Eqs. \eqref{D k,m_a n_a} and \eqref{phi}.

\subsection{Results}
We consider the violation of Ineq.~\eqref{CHSH_inequality} which uses two settings and two outcomes per party. While many strategies can be used to bin the measurement results, we could only find a violation for the parity binning, which corresponds to the measurement of  $(-1)^{ \hat{J}_{\theta_A,\phi_A}^A}$ for Alice and $(-1)^{ \hat{J}_{\theta_B,\phi_B}^B}$ for Bob. We present results obtained in this case below.\\

First, we fix the total number of spins $N$ and we optimize the value of~\eqref{CHSH_inequality} over the measurement settings for various values of $\chi t.$ The result is shown in Fig. \ref{fig2} for $N=6$ and $N=10.$ For low $\chi t,$ the violation increases until a value which depends on the number of atoms and then goes down whereas in the extreme squeezing regime the violation can go higher.\\

\begin{figure}[ht!]
\includegraphics[width = 7.5cm]{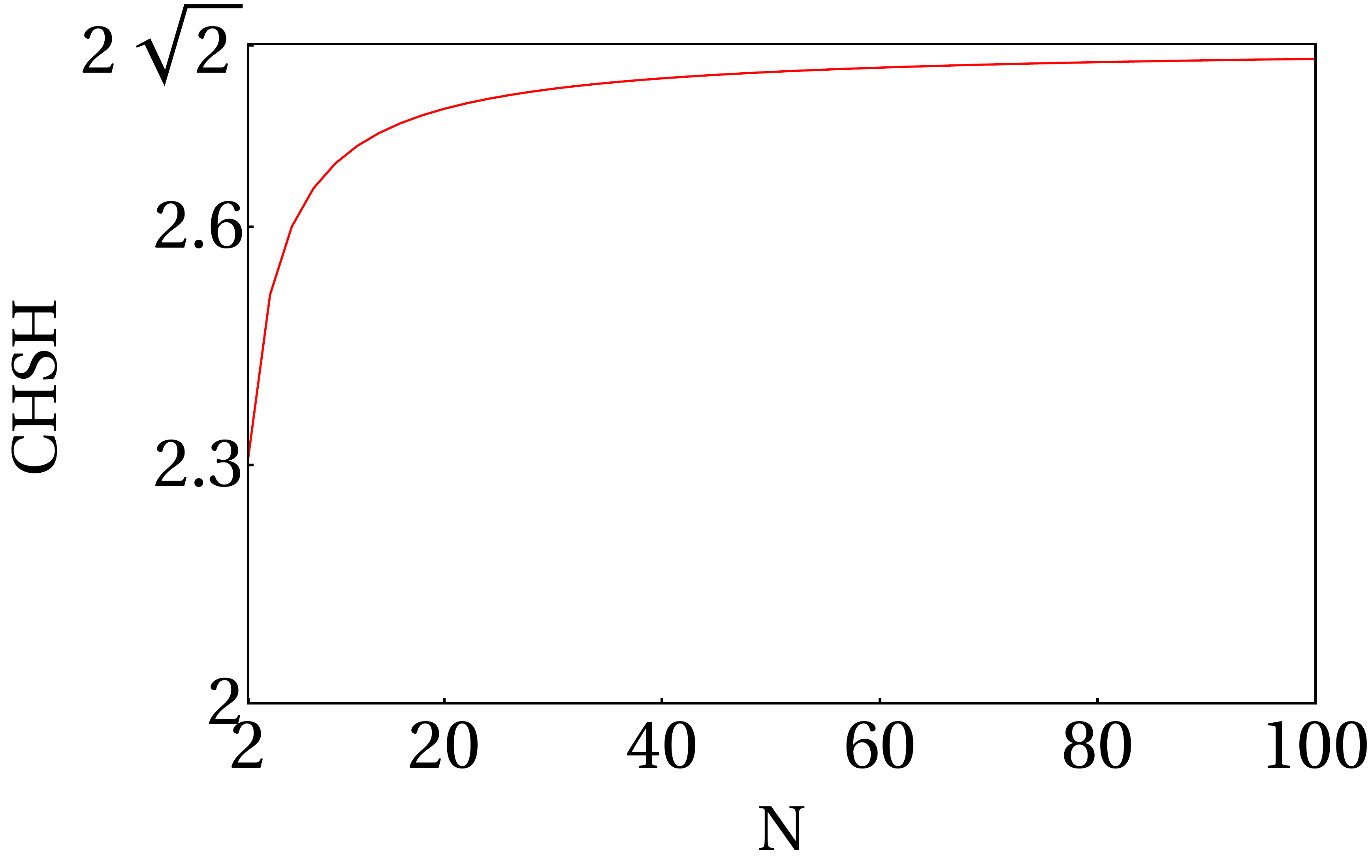}
\caption{Violation of the inequality \eqref{CHSH_inequality} for a spin squeezed state $\ket{\phi}$ with $\chi t =\frac{\pi}{2}$ as a function of the total number of atoms $N.$}
\label{fig3}
\end{figure}

We further investigate the extreme squeezing case where $\chi t = \pi/2.$ Remarkably the violation increases with the atom number and seems to saturate very close to the maximum value achievable by quantum states $2\sqrt{2}$, see Fig. \ref{fig3}. This implies that it is possible to self test a singlet state and Pauli measurements with collective observables. Although this result is unexpected, it is extremely challenging if not completely out of reach experimentally as the maximally squeezed state corresponds to a GHZ state when $\chi t = \pi/2.$  \\

\begin{figure}[ht!]
\includegraphics[width = 7.5cm]{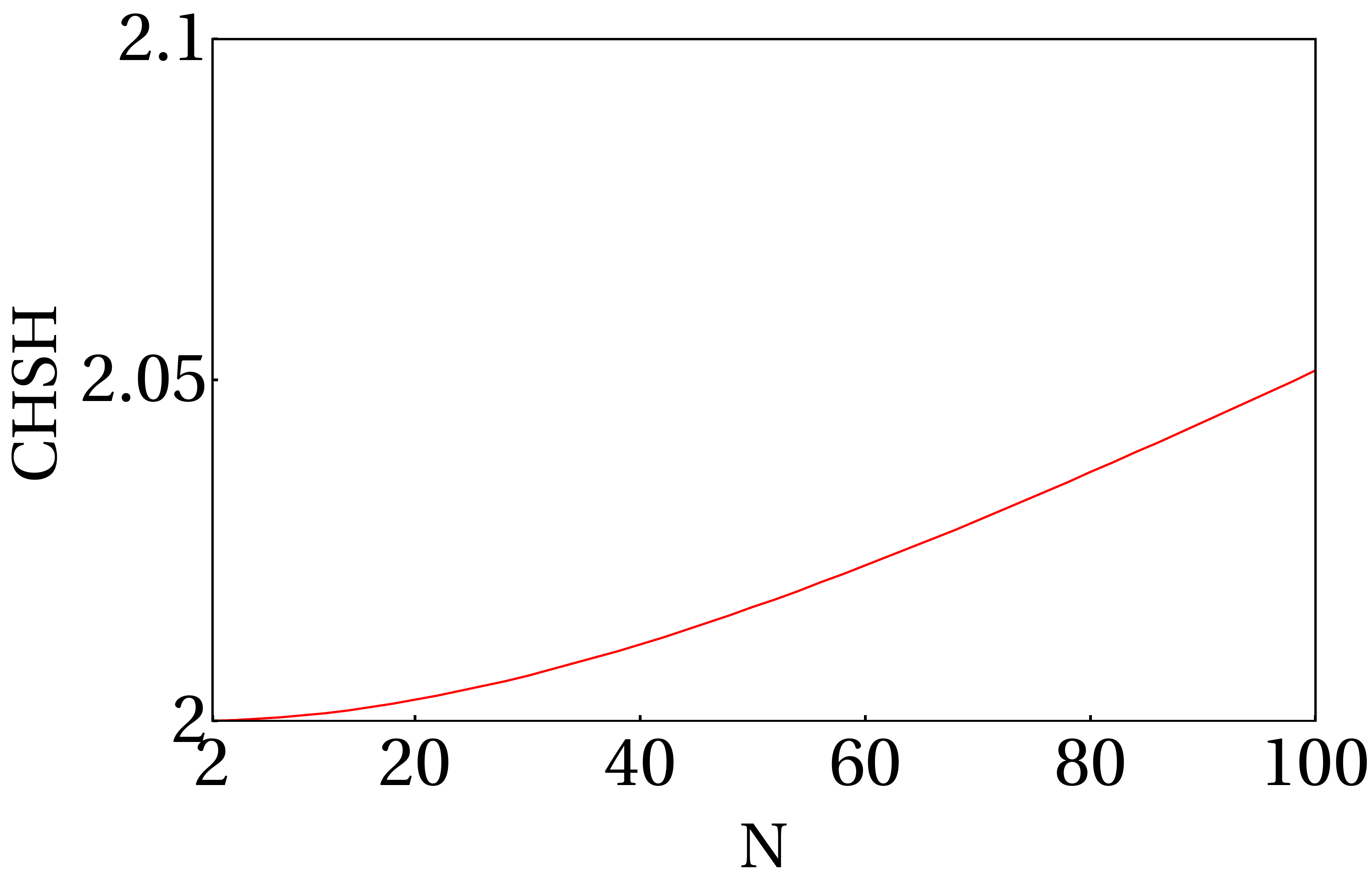}
\caption{Violation of CHSH for a small $\chi t=0.006$ which correspond to 10 dB of squeezing at 500 atoms. The violation is plotted with respect to the total atom number.}
\label{fig4}
\end{figure}

We thus focus on the regime where $\chi t$ is small, which is the most relevant regime in practice. We fix $\chi t = 0.006$
corresponding to a Wineland squeezing parameter $10\log{\xi^2}=10$ dB for $N=500$ atoms where $\xi^2=\frac{N(\Delta \hat{J}_\bot^A)^2}{\langle \hat{J}_x^A\rangle^2}$ \cite{Wineland92, Wineland94}, $\hat{J}_\bot^A$ corresponding a projection along the squeezing direction (before splitting). The resulting violation increases as a function of the atom number, as shown in Fig. \ref{fig4}.\\

\section{Conclusion}
\label{section_4}
We investigated the possibility of detecting bipartite nonlocality in many-body systems. We devoted a special attention to the experimental realization by considering realistic measurements. In particular, we focused on collective measurements only where spins are all measured in the same direction locally. We showed numerical results suggesting that no-Bell inequality with first order correlators can be violated whenever such measurements act of $N\geq 2$ spins. We then proved that the CHSH-Bell inequality can be violated with collective measurements as long as parity measurements can be performed. This suggests that parity measurements is a key ingredient -- even maybe necessary -- to reveal bipartite correlations in many-body systems with Bell inequalities.

\begin{acknowledgments}
This work was supported by the Swiss National Science Foundation (SNSF) through the Grant PP00P2-179109. We also acknowledge the Army Research Laboratory Center for Distributed Quantum Information via the project SciNet.\\
\end{acknowledgments}

\bibliographystyle{plain}

\appendix
\clearpage

\section{Complete list of facets for the local polytope in the Bell scenario $[(2, 2, 2, 2), (2, 2, 2, 2)]$}
The local polytope for two parties and four binary settings per party was recently solved in~\cite{Deza16}. This work demonstrates that the polytope admits 175 classes of inequalities, up to permutations of parties, inputs and outputs, but it does not provide a description of these inequalities. Still, some of them can be found in the litterature. First, the positivity constraint, stating that probabilities are larger than 0, is known to be a (trivial) facet of every local polytope. Nontrivial Bell inequalities involving fewer than 4 settings for each party also give rise to lifted Bell inequalities in this scenario. This includes the CHSH inequality~\cite{CHSH69}, $I_{3322}$~\cite{Froissart81}, as well as the three $I_{4322}^{1,2,3}$ inequalities~\cite{Collins04}.

But several inequalities using four measurements settings per party were also already known. Such is the case for $A_5$, $A_6$, $A_7$, $AS_1$, $AS_2$, $AII_1$, $AII_2$ and the $I_{4422}^i$ inequalities with $i\in\{1,\ldots,20\}$~\cite{Brunner08, Ito06, Gisin07, OpenProblems, Collins04, Avis05}, as well as the $J_{4422}^i$ inequalities with $i\in\{1,\ldots,129\}$ from~\cite{Pal09} and the symmetric $S_{242}^i$ inequalities with $i\in\{1,\ldots,52\}$ from~\cite{Bancal10}. Some of these inequalities are equivalent to each other, thus forming altogether 163 distinct families.

Noticing that randomly changing a few coefficients of a Bell inequality typically results in a non-facet defining Bell inequality with high rank, we used the algorithm given in section 2.3 of~\cite{Bancal10} to find tight facets below such inequalities for a number of random modifications of these 163 inequalities. This allowed us to discover a few additional inequalities, which were completed by a list of inequalities found independently by N. Brunner~\cite{Brunner09}. We refer to these additional 12 inequalities as $N_{4422}^i$ with $i\in\{1,\ldots,12\}$. Altogether, this provides a complete list of 175 families of tight Bell inequalities, which we present here.

A Bell inequality for $m=4$ binary settings is defined by 25 parameters. Following the main text we use the correlation picture with outcomes $a_i, b_j\in\{-1,1\}$. A Bell inequality can then be written as 
\begin{eqnarray}
\nonumber
&B&=\delta +\sum_{i=1}^{4} \alpha_i \langle a_i \rangle +\sum_{j=1}^{4} \beta_j \langle b_j \rangle + \sum_{i,j=1}^{4} \gamma_{ij} \langle a_i b_j\rangle \geq 0
\end{eqnarray}
or equivalently, in table format
\begin{equation}
B=\left(\begin{array}{c|cccc}
\ell & \beta_1 & \beta_2 & \beta_3 & \beta_4\\
\hline
\alpha_1 & \gamma_{11} & \gamma_{12} & \gamma_{13} & \gamma_{14}\\
\alpha_2 & \gamma_{21} & \gamma_{22} & \gamma_{23} & \gamma_{24}\\
\alpha_3 & \gamma_{31} & \gamma_{32} & \gamma_{33} & \gamma_{34}\\
\alpha_4 & \gamma_{41} & \gamma_{42} & \gamma_{43} & \gamma_{44}\\
\end{array}\right)\geq 0,
\end{equation}
In the main text we used $v_{a_i}'=-\alpha_i$, $v_{b_j}'=-\beta_j$, $w_{ij}'=-\gamma_{ij}$ and $\ell=\delta$.

The coefficients of the 175 classes of inequalities are given in table~\ref{I4422} together with the name under which they have been referred to earlier in the litterature. The are also provided in electronic format online~\cite{ancillary}. The first 6 inequalities are liftings from simpler Bell scenarios, after which come 169 inequalities which truly involve all four settings of both parties.

\onecolumngrid
%\clearpage
%\newpage

\renewcommand*{\arraystretch}{1.4}

\begin{longtable}{|c|@{\;}r@{\;}r@{\;}r@{\;}r@{\;}r|@{\;}r@{\;}r@{\;}r@{\;}r@{\;}r|@{\;}r@{\;}r@{\;}r@{\;}r@{\;}r|@{\;}r@{\;}r@{\;}r@{\;}r@{\;}r|@{\;}r@{\;}r@{\;}r@{\;}r@{\;}r@{\;}|c|}
\caption{List of $I_{4422}$ inequalities}\label{I4422}\\
  \hline   &&&&&&&&&&&&&&&&&&&&&&&&&&\\[.01em]
  \#&$\delta$&$\alpha_1$&$\alpha_2$&$\alpha_3$&$\alpha_4$&$\beta_1$&$\gamma_{11}$&$\gamma_{21}$&$\gamma_{31}$&$\gamma_{41}$&$\beta_2$&$\gamma_{12}$&$\gamma_{22}$&$\gamma_{32}$&$\gamma_{42}$&$\beta_3$&$\gamma_{13}$&$\gamma_{23}$&$\gamma_{33}$&$\gamma_{43}$&$\beta_4$&$\gamma_{14}$&$\gamma_{24}$&$\gamma_{34}$&$\gamma_{44}$&short name\\[.2em]\hline
  &&&&&&&&&&&&&&&&&&&&&&&&&&\\[.01em]
\endfirsthead
\caption{continued}\\
  \hline  &&&&&&&&&&&&&&&&&&&&&&&&&&\\[.01em]
\#&$\delta$&$\alpha_1$&$\alpha_2$&$\alpha_3$&$\alpha_4$&$\beta_1$&$\gamma_{11}$&$\gamma_{21}$&$\gamma_{31}$&$\gamma_{41}$&$\beta_2$&$\gamma_{12}$&$\gamma_{22}$&$\gamma_{32}$&$\gamma_{42}$&$\beta_3$&$\gamma_{13}$&$\gamma_{23}$&$\gamma_{33}$&$\gamma_{43}$&$\beta_4$&$\gamma_{14}$&$\gamma_{24}$&$\gamma_{34}$&$\gamma_{44}$&short name\\[.2em]\hline
  &&&&&&&&&&&&&&&&&&&&&&&&&&\\[.01em]
\endhead
  \multicolumn{13}{l}{{Continued on the next page\ldots}} \\
\endfoot
  \hline
\endlastfoot
1&1&1&0&0&0&1&1&0&0&0&0&0&0&0&0&0&0&0&0&0&0&0&0&0&0&positivity\\[.01em]
2&2&0&0&0&0&0&1&1&0&0&0&1&-1&0&0&0&0&0&0&0&0&0&0&0&0&CHSH\\[.01em]
3&4&1&1&0&0&1&1&1&1&0&1&1&1&-1&0&0&1&-1&0&0&0&0&0&0&0&$I_{3322}$\\[.01em]
4&5&1&1&1&0&1&1&1&1&0&0&1&0&-1&0&0&1&-1&0&0&0&0&1&-1&0&$I_{4322}^2$\\[.01em]
5&6&0&0&0&0&1&1&1&1&0&1&-1&-1&-1&0&0&1&1&-2&0&0&1&-1&0&0&$I_{4322}^3$\\[.01em]
6&6&2&0&0&0&1&1&1&1&0&1&1&1&-1&0&1&1&-1&1&0&1&1&-1&-1&0&$I_{4322}^1$\\[.01em]
7&6&0&0&0&0&0&1&1&1&1&0&1&1&1&-1&0&1&1&-2&0&0&1&-1&0&0&$AS_1$\\[.01em]
8&6&1&1&0&0&0&1&-1&1&1&0&1&-1&1&-1&0&1&0&-1&0&0&0&1&1&0&$I_{4422}^5$\\[.01em]
9&6&1&1&0&0&1&1&1&1&0&1&1&0&-1&1&0&1&-1&-1&-1&0&0&1&-1&0&$A_5$\\[.01em]
10&6&2&0&0&0&1&1&1&1&0&1&1&1&-1&0&1&1&-1&0&1&1&1&-1&0&-1&$I_{4422}^3$\\[.01em]
11&7&1&1&1&0&1&1&1&0&1&1&1&0&1&-1&1&-1&1&1&0&0&0&1&-1&-2&$I_{4422}^{16}$\\[.01em]
12&7&1&1&1&0&1&1&1&0&1&0&1&-1&-1&1&0&1&0&-1&-2&0&0&1&-1&0&$J_{4422}^{44}$\\[.01em]
13&7&1&1&1&0&1&1&1&1&0&0&1&1&-1&1&0&1&1&-1&-1&0&2&-2&0&0&$J_{4422}^{25}$\\[.01em]
14&7&1&1&1&0&1&1&1&1&2&1&1&1&0&-1&1&1&0&1&-1&0&0&1&-1&0&$I_{4422}^{7}$\\[.01em]
15&7&2&1&1&1&2&1&1&1&1&1&1&1&0&-1&1&1&0&-1&1&1&1&-1&1&0&$A_6$\\[.01em]
16&7&2&2&1&0&2&1&1&1&1&2&1&1&1&-1&1&1&1&-1&0&0&1&-1&0&0&$A_7$, $I_{4422}^1$\\[.01em]
17&8&1&1&0&0&1&1&0&1&1&1&0&1&-1&-1&0&1&-1&-3&1&0&1&-1&1&-1&$J_{4422}^{60}$\\[.01em]
18&8&1&1&0&0&1&1&1&1&2&1&1&0&1&-1&0&1&1&1&-1&0&2&-1&-1&0&$AII_1$\\[.01em]
19&8&1&1&0&0&1&2&1&1&1&1&0&1&-1&-1&0&2&-1&-1&0&0&1&0&1&-2&$J_{4422}^{43}$\\[.01em]
20&8&1&1&0&0&1&2&1&2&0&1&1&2&-2&0&0&1&-1&-1&1&0&1&-1&-1&-1&$I_{4422}^2$\\[.01em]
21&8&1&1&1&1&1&1&2&1&-1&1&2&-2&1&0&1&1&1&0&1&1&-1&0&1&1&$I_{4422}^{14}$\\[.01em]
22&8&1&1&1&1&1&2&1&-1&1&1&1&1&1&0&0&1&-1&1&-1&0&1&-2&0&1&$J_{4422}^{69}$\\[.01em]
23&8&1&1&1&1&1&2&1&1&-1&1&1&2&-1&1&0&1&-1&-1&1&0&1&-1&0&0&$AII_2$\\[.01em]
24&8&2&1&1&0&1&1&2&-1&1&1&1&1&0&-1&0&1&-1&-1&1&0&1&-1&-1&-1&$J_{4422}^{74}$\\[.01em]
25&8&2&1&1&0&2&1&1&1&1&1&1&-2&1&-1&1&1&1&0&-1&0&1&-1&-1&1&$I_{4422}^{15}$\\[.01em]
26&8&2&1&1&0&2&1&1&1&1&1&1&2&-1&-1&1&1&-1&0&-1&0&1&-1&-1&1&$I_{4422}^{13}$\\[.01em]
27&8&2&1&1&0&2&1&2&1&2&1&1&-1&1&0&1&1&1&0&-1&0&1&-1&-1&1&$J_{4422}^6$\\[.01em]
28&8&2&1&1&0&2&1&2&2&1&1&1&1&-1&0&1&1&-1&1&0&0&1&-1&-1&1&$J_{4422}^1$\\[.01em]
29&8&2&2&0&0&0&1&-1&1&1&0&1&-1&1&-1&0&1&-1&-1&1&0&1&-1&-1&-1&$I_{4422}^6$\\[.01em]
30&8&2&2&0&0&2&1&1&1&1&2&1&1&-1&-1&0&1&-1&1&-1&0&1&-1&-1&1&$I_{4422}^4$\\[.01em]
31&9&1&1&1&0&1&-3&2&1&1&1&2&1&1&1&1&1&1&0&-1&0&1&1&-1&1&$J_{4422}^{26}$\\[.01em]
32&9&1&1&1&0&1&-1&2&1&1&1&2&1&0&2&1&1&0&1&-1&0&1&2&-1&-2&$I_{4422}^{12}$\\[.01em]
33&9&1&1&1&0&1&1&1&1&0&0&2&-1&-2&1&0&1&-2&2&1&0&1&-1&0&-2&$J_{4422}^{63}$\\[.01em]
34&9&1&1&1&0&1&1&1&2&1&0&2&1&-2&1&0&1&-1&1&-1&0&1&-2&0&1&$J_{4422}^{75}$\\[.01em]
35&9&1&1&1&0&1&1&1&2&1&1&1&1&2&-1&1&2&2&-3&0&0&1&-1&0&0&$J_{4422}^{32}$\\[.01em]
36&9&1&1&1&0&1&2&1&1&1&1&-2&2&1&0&1&2&1&0&-2&0&1&1&-1&1&$J_{4422}^{23}$\\[.01em]
37&9&1&1&1&0&1&2&1&1&1&1&-1&1&2&1&1&1&1&0&-1&0&1&-2&2&-1&$J_{4422}^{66}$\\[.01em]
38&9&1&1&1&0&1&2&1&1&1&1&1&1&1&-2&1&-1&2&1&1&0&1&1&-2&0&$I_{4422}^{11}$\\[.01em]
39&9&1&1&1&0&1&2&2&-2&1&1&1&1&2&1&1&1&1&1&-2&0&1&-1&0&0&$J_{4422}^{81}$\\[.01em]
40&9&2&1&0&0&1&2&1&1&1&1&2&1&1&-1&1&2&1&-2&0&0&2&-2&0&0&$J_{4422}^{71}$\\[.01em]
41&9&2&1&0&0&1&2&1&1&1&1&1&1&1&-2&1&1&1&-1&0&0&2&-2&-1&-1&$J_{4422}^{88}$\\[.01em]
42&9&2&1&0&0&1&2&1&1&1&1&2&1&1&-1&1&1&1&-1&0&0&3&-2&-1&0&$J_{4422}^{33}$\\[.01em]
43&9&2&1&1&1&1&2&1&1&-1&1&2&1&-1&1&1&0&-1&1&1&0&2&-2&0&0&$J_{4422}^{54}$\\[.01em]
44&10&0&0&0&0&0&2&2&1&1&0&2&-1&-1&-2&0&1&-1&-2&2&0&1&-2&2&1&$AS_2$\\[.01em]
45&10&1&1&0&0&0&2&-1&2&1&0&1&-1&-2&2&0&1&-2&-1&-2&0&1&1&-1&-1&$J_{4422}^{111}$\\[.01em]
46&10&1&1&0&0&1&1&2&1&1&1&2&0&-2&-1&0&1&-2&1&2&0&1&-1&2&-2&$I_{4422}^{17}$\\[.01em]
47&10&2&0&0&0&1&2&1&1&1&1&2&1&0&-2&1&2&0&-2&1&1&2&-2&1&0&$J_{4422}^{118}$\\[.01em]
48&10&2&0&0&0&1&2&2&2&1&1&2&-1&-1&-1&1&1&1&-2&1&1&1&-2&1&1&$J_{4422}^{14}$\\[.01em]
49&10&2&1&1&0&2&1&2&2&1&1&2&-1&1&-1&1&2&1&-2&0&0&1&-1&0&2&$I_{4422}^8$\\[.01em]
50&10&2&1&1&0&2&2&2&1&1&1&1&-1&2&-1&1&2&0&-2&-1&0&1&-2&0&1&$J_{4422}^{37}$\\[.01em]
51&10&2&1&1&0&2&2&2&1&1&1&2&1&-1&-1&1&0&-1&1&-1&0&2&-3&0&1&$J_{4422}^{20}$\\[.01em]
52&10&2&2&0&0&2&2&2&1&1&2&1&1&-1&-1&0&2&-2&1&-1&0&1&1&-1&-1&$N_{4422}^{10}$\\[.01em]
53&10&2&2&1&1&2&1&1&1&1&0&2&-1&1&-2&0&2&-1&-2&1&0&1&-1&1&1&$N_{4422}^4$\\[.01em]
54&10&3&1&0&0&2&2&1&2&1&2&1&1&-1&-1&1&2&-1&1&-1&1&2&0&-2&1&$J_{4422}^{9}$\\[.01em]
55&10&3&1&1&1&3&2&2&2&1&1&2&-1&1&-1&1&2&1&-2&0&1&1&-1&0&1&$I_{4422}^9$\\[.01em]
56&11&1&1&1&0&1&2&2&-1&2&0&1&1&-2&-4&0&1&-1&1&-1&0&1&-1&-1&1&$J_{4422}^{82}$\\[.01em]
57&11&2&1&0&0&1&2&-1&2&2&1&2&-1&1&-3&1&1&1&-1&0&0&1&-2&-2&1&$J_{4422}^{67}$\\[.01em]
58&11&2&1&1&1&1&3&-2&1&1&1&1&1&-1&2&1&1&1&1&0&0&3&1&-2&-2&$J_{4422}^{51}$\\[.01em]
59&11&2&1&1&1&1&3&-2&1&1&1&2&1&-1&1&1&1&1&2&-1&0&2&1&-1&-2&$N_{4422}^{7}$\\[.01em]
60&11&2&1&1&1&2&1&2&2&1&1&2&2&-1&-2&1&1&1&-1&2&1&2&-2&1&0&$J_{4422}^{2}$\\[.01em]
61&11&2&1&1&1&2&1&2&2&1&1&2&-3&1&1&1&2&1&-1&-1&1&1&1&-1&2&$J_{4422}^{102}$\\[.01em]
62&11&2&1&1&1&2&2&1&1&2&1&3&1&1&-2&0&2&-2&-1&1&0&1&1&-2&0&$J_{4422}^{87}$\\[.01em]
63&11&2&1&1&1&2&2&1&1&2&1&3&1&1&-2&0&3&-2&-2&1&0&0&1&-1&0&$J_{4422}^{83}$\\[.01em]
64&11&2&1&1&1&2&2&2&1&1&1&1&-2&2&0&0&2&-1&-1&-2&0&1&-2&-1&2&$J_{4422}^{112}$\\[.01em]
65&11&2&1&1&1&2&2&2&1&1&1&1&-1&2&-1&0&2&-2&-1&-1&0&1&2&-1&-2&$J_{4422}^{94}$\\[.01em]
66&11&2&1&1&1&2&2&2&2&0&1&2&1&-1&1&0&3&-3&1&-1&0&1&1&-1&-1&$J_{4422}^{24}$\\[.01em]
67&11&2&2&1&0&2&2&1&1&2&2&1&1&2&-2&1&1&2&-1&-1&0&2&-2&-1&-1&$I_{4422}^{10}$\\[.01em]
68&11&3&1&1&0&2&2&1&2&1&1&3&-2&1&-1&1&3&1&-2&1&1&1&1&0&-1&$J_{4422}^{35}$\\[.01em]
69&11&3&1&1&0&2&2&2&2&0&1&2&1&-1&1&1&2&1&-1&-1&1&3&-3&1&0&$J_{4422}^{36}$\\[.01em]
70&12&1&1&0&0&1&-1&2&2&2&1&2&1&-1&-1&0&2&-1&-2&3&0&2&-1&3&0&$J_{4422}^{22}$\\[.01em]
71&12&1&1&0&0&1&-1&2&2&2&1&2&1&-1&-1&0&2&-1&4&-1&0&2&-1&-1&2&$J_{4422}^{61}$\\[.01em]
72&12&1&1&0&0&1&3&1&2&1&1&1&3&-2&-1&0&2&-2&-2&-2&0&1&-1&-2&2&$J_{4422}^{27}$\\[.01em]
73&12&1&1&1&1&1&-2&1&3&1&1&2&2&1&0&0&2&-2&2&-2&0&1&-2&1&2&$J_{4422}^{72}$\\[.01em]
74&12&1&1&1&1&1&2&-1&-2&2&1&-2&2&1&2&1&1&-2&3&1&1&2&2&1&0&$J_{4422}^{41}$\\[.01em]
75&12&1&1&1&1&1&-2&3&1&1&1&2&1&1&1&0&2&2&-3&-1&0&1&1&2&-2&$J_{4422}^{76}$\\[.01em]
76&12&1&1&1&1&1&-1&2&-2&2&1&2&3&1&-1&1&-2&1&3&1&1&2&-1&1&1&$J_{4422}^{62}$\\[.01em]
77&12&1&1&1&1&1&2&-3&3&1&1&1&2&1&1&0&2&-1&-2&1&0&2&1&1&-2&$J_{4422}^{106}$\\[.01em]
78&12&1&1&1&1&1&2&3&1&-1&1&-2&1&3&1&0&2&-2&2&-2&0&1&-1&1&1&$J_{4422}^{126}$\\[.01em]
79&12&1&1&1&1&1&2&3&2&-2&1&1&1&1&2&0&2&-2&1&-1&0&2&1&-3&0&$J_{4422}^{77}$\\[.01em]
80&12&2&1&1&0&1&1&-2&3&1&1&2&1&1&-1&0&2&-2&-2&2&0&1&-2&-1&-2&$J_{4422}^{116}$\\[.01em]
81&12&2&1&1&0&1&2&2&-1&2&1&2&1&1&-1&0&3&-3&-1&1&0&1&1&-2&-2&$J_{4422}^{50}$\\[.01em]
82&12&2&1&1&0&2&0&2&2&2&1&2&-1&2&-2&1&2&1&-1&-1&0&2&-1&0&3&$J_{4422}^{16}$\\[.01em]
83&12&2&1&1&0&2&0&2&2&2&1&2&1&-1&-1&1&2&-1&1&-1&0&2&-1&-1&4&$J_{4422}^{19}$\\[.01em]
84&12&2&1&1&0&2&1&2&3&2&1&2&2&-2&1&1&2&-1&1&-1&0&1&-2&-1&2&$J_{4422}^{4}$\\[.01em]
85&12&2&1&1&0&2&2&2&1&1&1&2&-1&-2&-2&1&1&-2&3&-1&0&1&-2&-1&2&$J_{4422}^{42}$\\[.01em]
86&12&2&1&1&0&2&2&3&1&2&1&3&-2&1&1&1&1&1&1&-2&0&2&1&-2&-1&$J_{4422}^{90}$\\[.01em]
87&12&2&1&1&0&2&2&3&2&1&1&3&-4&1&1&1&2&1&-1&-1&0&1&1&-1&1&$J_{4422}^{58}$\\[.01em]
88&12&2&2&1&1&2&2&2&1&-1&2&2&0&-2&2&1&1&-2&3&1&1&-1&2&1&1&$J_{4422}^{17}$\\[.01em]
89&12&3&1&0&0&2&2&1&2&1&2&1&2&-2&-1&1&2&-1&1&-3&1&2&-1&-1&1&$J_{4422}^{34}$\\[.01em]
90&12&3&1&0&0&2&3&1&3&1&2&2&2&-2&0&1&2&-1&-1&1&1&2&-1&0&-2&$J_{4422}^{121}$\\[.01em]
91&12&3&1&1&1&2&2&3&1&-2&2&2&1&-1&2&2&2&-2&2&0&0&1&-1&-1&-1&$J_{4422}^{59}$\\[.01em]
92&12&3&1&1&1&3&1&2&2&2&1&2&-2&-1&2&1&2&-1&1&-1&1&2&2&-1&0&$I_{4422}^{19}$\\[.01em]
93&12&3&2&1&0&3&2&2&1&2&2&1&2&-1&-2&1&2&-1&1&-1&0&2&-1&-2&1&$J_{4422}^{31}$\\[.01em]
94&12&4&1&1&0&3&2&2&1&2&2&2&-2&1&-1&2&2&2&0&-2&1&2&-1&-1&1&$J_{4422}^{7}$\\[.01em]
95&12&4&1&1&0&3&2&2&1&2&2&2&-1&2&-1&2&2&1&-1&-2&1&2&-1&-1&1&$J_{4422}^{8}$\\[.01em]
96&12&4&1&1&0&3&2&2&2&1&2&2&2&-1&-1&2&2&-2&1&-1&1&2&-1&-1&1&$J_{4422}^{30}$\\[.01em]
97&12&4&2&0&0&3&2&2&2&1&3&2&2&-2&-1&1&2&-1&1&-1&1&2&-1&-1&1&$J_{4422}^{11}$\\[.01em]
98&12&4&2&1&1&4&2&2&2&2&2&2&2&-1&-1&1&2&-1&1&-1&1&2&-1&-1&1&$I_{4422}^{20}$\\[.01em]
99&13&1&1&1&0&1&-3&2&3&1&1&2&-1&2&-2&1&3&2&1&1&0&1&-2&1&2&$J_{4422}^{110}$\\[.01em]
100&13&1&1&1&0&1&2&1&2&2&1&-2&3&1&-1&1&3&1&0&-3&0&2&2&-2&2&$J_{4422}^{70}$\\[.01em]
101&13&1&1&1&0&1&2&3&-2&2&1&2&1&1&-3&1&1&1&3&2&0&2&-2&-1&1&$J_{4422}^{93}$\\[.01em]
102&13&2&1&0&0&1&2&2&2&1&1&2&1&-2&2&1&3&1&-1&-2&0&3&-3&1&1&$J_{4422}^{107}$\\[.01em]
103&13&2&1&1&1&1&2&2&-1&2&1&-1&-2&2&2&1&2&1&1&-1&0&3&-4&-1&0&$J_{4422}^{45}$\\[.01em]
104&13&2&1&1&1&1&2&2&2&-1&1&3&-2&1&1&1&0&2&-1&2&0&3&1&-3&-1&$J_{4422}^{47}$\\[.01em]
105&13&2&1&1&1&2&0&2&2&2&1&2&-3&3&-1&1&2&3&1&-1&1&2&-1&-1&1&$J_{4422}^{28}$\\[.01em]
106&13&2&1&1&1&2&1&2&1&2&1&2&-3&3&-1&1&1&3&1&-2&1&2&-1&-2&0&$J_{4422}^{86}$\\[.01em]
107&13&2&2&1&0&2&3&2&2&1&1&1&2&-1&-1&0&2&1&-2&1&0&4&-3&0&-1&$J_{4422}^{48}$\\[.01em]
108&13&3&1&1&0&2&2&-3&2&1&2&2&3&1&2&2&1&2&1&-2&1&2&-1&-1&-1&$J_{4422}^{15}$\\[.01em]
109&13&3&1&1&0&2&2&3&-1&2&2&2&1&1&-2&1&2&-2&-1&2&0&1&-1&-2&-2&$J_{4422}^{122}$\\[.01em]
110&13&3&2&1&1&3&2&2&2&1&1&2&2&-1&-2&1&2&1&-1&1&0&3&-3&1&-1&$N_{4422}^{3}$\\[.01em]
111&13&4&1&0&0&2&3&1&2&2&2&2&1&1&-2&2&2&1&-2&1&1&3&-2&-1&-1&$N_{4422}^{2}$\\[.01em]
112&14&2&1&1&0&1&2&2&-1&2&1&3&1&1&-2&0&3&-3&-2&2&0&0&1&-3&-2&$J_{4422}^{129}$\\[.01em]
113&14&2&1&1&0&1&2&3&-1&3&1&-1&2&2&-2&1&3&1&-1&-4&1&2&-1&1&1&$J_{4422}^{80}$\\[.01em]
114&14&2&1&1&0&1&3&1&-1&4&1&2&1&1&-1&0&2&2&-2&-2&0&3&-3&-1&-1&$J_{4422}^{64}$\\[.01em]
115&14&2&1&1&0&1&3&2&-1&3&1&2&1&1&-1&0&2&2&-2&-2&0&3&-4&-1&0&$J_{4422}^{68}$\\[.01em]
116&14&2&1&1&0&2&2&2&3&1&1&2&-2&-1&-2&1&1&3&-3&0&0&1&-2&-2&3&$J_{4422}^{78}$\\[.01em]
117&14&2&1&1&0&2&3&3&2&2&1&3&-2&1&1&1&2&1&1&-3&0&2&1&-3&0&$S_{242}^{51}$\\[.01em]
118&14&2&2&1&1&2&2&1&2&1&1&2&3&-2&-2&1&1&1&-2&3&0&3&-3&-1&-1&$J_{4422}^{99}$\\[.01em]
119&14&2&2&1&1&2&3&2&2&-1&2&2&0&2&2&1&2&2&-4&1&1&-1&2&1&-1&$S_{242}^{52}$\\[.01em]
120&14&2&2&1&1&2&3&2&2&-1&2&2&3&-1&2&0&4&-4&-1&1&0&1&1&-1&-1&$J_{4422}^{124}$\\[.01em]
121&14&3&2&1&0&3&1&3&3&2&2&2&2&-2&0&1&2&-1&1&-1&0&2&-2&-1&3&$J_{4422}^{21}$\\[.01em]
122&14&4&1&1&0&3&3&3&1&2&2&3&1&-1&-3&2&2&-2&2&0&1&2&-1&-1&1&$J_{4422}^{127}$\\[.01em]
123&14&4&2&1&1&4&2&2&2&2&2&2&1&-2&-1&1&2&-2&-1&2&1&2&-1&2&-2&$I_{4422}^{18}$\\[.01em]
124&14&4&2&1&1&4&2&3&3&2&2&2&2&-2&0&1&2&-2&-1&2&1&2&-1&1&-1&$J_{4422}^{5}$\\[.01em]
125&14&4&2&2&0&3&2&2&1&2&3&2&1&2&-2&1&3&1&-2&-1&1&3&-2&1&1&$N_{4422}^{6}$\\[.01em]
126&15&2&1&1&1&1&2&2&2&-1&1&4&-3&1&1&1&1&2&-1&3&0&3&2&-3&-2&$J_{4422}^{46}$\\[.01em]
127&15&2&1&1&1&2&1&3&3&1&1&3&-2&2&-2&1&3&2&-3&1&1&1&-2&1&3&$J_{4422}^{108}$\\[.01em]
128&15&2&1&1&1&2&2&3&2&1&1&4&-3&1&1&0&3&2&-2&-3&0&1&1&-2&2&$J_{4422}^{89}$\\[.01em]
129&15&2&1&1&1&2&3&2&2&1&1&1&-2&-1&3&0&3&-2&-3&-2&0&1&-3&3&-1&$J_{4422}^{96}$\\[.01em]
130&15&2&1&1&1&2&3&3&1&1&1&3&-2&2&2&0&3&-1&1&-3&0&3&-1&-3&1&$J_{4422}^{117}$\\[.01em]
131&15&2&2&1&0&2&2&3&2&1&1&2&0&-1&-2&0&3&-4&-1&2&0&1&3&-3&1&$J_{4422}^{39}$\\[.01em]
132&15&2&2&2&1&2&2&2&3&-1&1&-3&-1&3&2&1&4&-3&1&1&1&1&2&-1&1&$J_{4422}^{53}$\\[.01em]
133&15&3&2&1&1&2&3&2&-1&2&2&2&3&2&-1&1&4&-4&1&0&0&2&1&-1&-2&$J_{4422}^{57}$\\[.01em]
134&15&3&2&1&1&2&3&2&2&-1&2&0&2&-2&2&1&3&-2&1&1&0&3&0&-4&-1&$J_{4422}^{55}$\\[.01em]
135&15&3&2&1&1&2&3&2&2&-1&2&1&2&-1&2&1&4&-3&1&1&0&3&1&-3&-1&$J_{4422}^{56}$\\[.01em]
136&15&3&2&1&1&3&1&3&3&2&2&2&3&-2&1&1&2&-3&-1&3&1&2&-1&1&-1&$J_{4422}^{3}$\\[.01em]
137&15&3&2&2&0&3&3&2&3&1&1&4&-3&1&-1&1&1&2&-1&-1&0&3&1&-3&1&$J_{4422}^{49}$\\[.01em]
138&15&4&2&1&0&3&2&2&2&1&2&3&2&-1&-2&1&4&-3&1&-1&1&3&1&-1&2&$N_{4422}^{9}$\\[.01em]
139&16&1&1&1&1&1&2&-3&2&4&1&2&3&1&1&0&2&-2&3&-3&0&3&-1&-3&-1&$J_{4422}^{98}$\\[.01em]
140&16&2&1&1&0&2&2&3&2&1&1&3&-4&-1&3&1&2&-1&-1&-3&0&1&3&-3&1&$J_{4422}^{85}$\\[.01em]
141&16&2&2&1&1&2&2&4&2&-2&2&1&2&2&3&1&3&-4&3&-1&1&2&0&-2&1&$J_{4422}^{79}$\\[.01em]
142&16&2&2&1&1&2&2&4&2&-2&2&2&1&-1&2&1&3&-4&3&-1&1&-1&1&3&2&$J_{4422}^{119}$\\[.01em]
143&16&2&2&1&1&2&2&3&4&-1&2&3&-1&2&2&1&4&2&-4&-1&1&-1&2&-1&1&$J_{4422}^{125}$\\[.01em]
144&16&2&2&1&1&2&4&2&2&-2&2&2&1&2&3&1&2&2&-4&1&1&-2&3&1&-1&$J_{4422}^{105}$\\[.01em]
145&16&3&1&1&1&2&3&3&3&-1&1&4&-5&1&-1&1&2&1&-1&1&0&2&2&-2&-2&$J_{4422}^{65}$\\[.01em]
146&16&3&1&1&1&3&3&2&3&1&1&2&-2&-2&3&1&3&-2&-1&-3&1&1&3&-3&0&$J_{4422}^{113}$\\[.01em]
147&16&3&2&1&0&2&4&2&-2&2&2&2&1&3&2&2&1&2&1&-2&0&4&-3&1&-2&$J_{4422}^{101}$\\[.01em]
148&16&4&1&1&0&2&3&2&1&2&2&3&1&2&-2&1&4&1&-3&-1&1&4&-3&1&1&$J_{4422}^{128}$\\[.01em]
149&16&5&1&1&1&3&3&3&1&-2&3&3&1&-2&3&3&3&-2&3&1&1&2&-1&-1&-1&$N_{4422}^{11}$\\[.01em]
150&16&5&2&2&1&5&3&3&2&3&2&3&-2&2&-1&2&2&2&0&-2&1&3&-1&-2&1&$J_{4422}^{13}$\\[.01em]
151&17&1&1&1&0&1&-3&2&4&2&1&2&3&1&-3&1&4&1&1&3&0&2&-3&3&-2&$J_{4422}^{91}$\\[.01em]
152&17&2&1&1&1&2&2&2&3&1&1&2&-3&-2&4&1&3&-2&-1&-3&1&1&4&-3&1&$J_{4422}^{92}$\\[.01em]
153&17&2&2&1&0&2&3&-1&3&3&1&1&2&-1&-1&0&2&-2&-2&2&0&2&-3&1&-6&$J_{4422}^{52}$\\[.01em]
154&17&3&1&1&0&2&2&-2&4&2&2&3&2&1&-2&1&3&-2&-3&3&0&1&-3&-1&-3&$J_{4422}^{115}$\\[.01em]
155&17&3&2&2&0&3&3&3&2&1&1&4&2&-3&-2&1&0&-1&2&-2&0&4&-4&1&1&$J_{4422}^{38}$\\[.01em]
156&18&2&2&1&1&2&-1&3&-3&3&2&3&2&2&-1&1&-3&1&5&2&1&3&-2&1&3&$J_{4422}^{40}$\\[.01em]
157&18&3&1&0&0&2&3&-1&4&4&2&3&-1&2&-4&2&2&2&-3&1&0&1&-3&-3&1&$J_{4422}^{84}$\\[.01em]
158&18&3&1&1&1&2&2&2&4&-2&2&3&4&-4&-1&2&1&2&2&3&2&3&-3&-1&1&$J_{4422}^{95}$\\[.01em]
159&18&3&2&1&0&3&3&2&3&1&2&1&4&-4&-1&1&3&-2&-1&-3&0&2&-2&-3&3&$J_{4422}^{120}$\\[.01em]
160&18&3&2&2&1&3&0&3&3&-3&2&3&2&-1&2&2&3&-2&2&1&1&-3&1&2&5&$J_{4422}^{18}$\\[.01em]
161&18&3&2&2&1&3&1&2&3&3&2&2&2&4&-2&2&3&4&-4&-1&1&3&-2&-1&1&$J_{4422}^{29}$\\[.01em]
162&18&3&2&2&1&3&3&2&1&3&2&4&2&2&-2&1&-5&4&3&-1&0&1&2&-2&-1&$J_{4422}^{114}$\\[.01em]
163&18&4&2&1&1&2&3&2&2&-1&2&2&2&-1&3&1&4&2&-3&-2&1&5&-4&1&1&$N_{4422}^{8}$\\[.01em]
164&18&4&2&1&1&3&2&3&3&1&2&4&2&-2&-2&2&2&1&-2&3&1&4&-4&2&1&$J_{4422}^{103}$\\[.01em]
165&18&4&2&1&1&3&3&2&3&1&2&4&2&-2&2&1&3&2&-1&-3&0&4&-4&1&-1&$N_{4422}^{1}$\\[.01em]
166&18&5&2&1&0&3&3&2&3&1&2&4&2&-2&2&2&3&2&-1&-2&1&5&-4&1&-1&$J_{4422}^{123}$\\[.01em]
167&18&6&2&1&1&5&3&3&3&2&3&3&2&-3&-1&2&3&-2&-1&2&2&3&-1&2&-2&$J_{4422}^{10}$\\[.01em]
168&18&6&2&2&2&6&3&3&3&3&2&3&2&-1&-2&2&3&-1&-2&2&2&3&-2&2&-1&$J_{4422}^{12}$\\[.01em]
169&19&2&2&1&0&2&3&2&1&2&1&-2&3&-3&-3&1&4&-1&2&-4&1&-3&2&5&-1&$J_{4422}^{73}$\\[.01em]
170&19&2&2&2&1&2&5&2&2&-3&2&2&2&1&3&1&-3&2&4&-2&0&2&-4&3&1&$J_{4422}^{100}$\\[.01em]
171&20&2&2&1&1&2&4&2&5&-1&1&3&2&-2&2&1&-4&5&1&1&0&3&3&-3&-3&$N_{4422}^{12}$\\[.01em]
172&20&3&2&1&0&3&2&3&3&1&2&3&-1&-2&-4&1&3&-2&-4&4&0&1&-4&4&1&$J_{4422}^{97}$\\[.01em]
173&21&3&2&1&1&3&4&5&3&-3&2&2&2&2&4&1&3&2&-5&1&1&4&-3&1&-1&$N_{4422}^{5}$\\[.01em]
174&23&2&2&2&1&2&2&6&2&-4&2&3&2&1&4&1&3&-4&5&-3&0&4&-2&-4&-2&$J_{4422}^{104}$\\[.01em]
175&24&2&2&1&1&2&6&2&-4&2&2&2&1&5&4&1&-4&5&-3&3&1&2&4&3&-4&$J_{4422}^{109}$\\[.01em]
\hline
\end{longtable}

\end{document}